\documentclass[twocolumn,secnumarabic,floatfix,amssymb,amsmath,nofootinbib,tightenlines,showpacs,superscriptaddress,aps,prl]{revtex4}
\usepackage{amsfonts}
\usepackage{bm}%
\usepackage[colorlinks=true,linkcolor=blue,pagecolor=blue,filecolor=blue,menucolor=yellow,urlcolor=blue,citecolor=blue,anchorcolor=blue]{hyperref}%
\usepackage{graphicx}
\usepackage{times}

\newcommand{\ie}{\textit{i.e. }}
\newcommand{\eg}{\textit{e.g. }}
\newcommand{\etal}{\emph{et al.}}
\newcommand{\sns}{S/N/S }
\newcommand{\sfs}{S/F/S }

\begin{document}

\title{Non-Fraunhofer Interference Pattern in Inhomogeneous Ferromagnetic Josephson Junctions }

\author{Mohammad Alidoust }
\email{phymalidoust@gmail.com} \affiliation{Department of Physics,
Norwegian University of Science and Technology, N-7491 Trondheim,
Norway}
\author{Granville Sewell}
\email{sewell@utep.edu} \affiliation{Mathematics Department,
University of Texas El Paso, El Paso, TX 79968, USA}
\author{Jacob Linder}
\email{jacob.linder@ntnu.no} \affiliation{Department of Physics,
Norwegian University of Science and Technology, N-7491 Trondheim,
Norway}
\date{\today}

\begin{abstract}
Generic conditions are established for producing a non-Fraunhofer
response of the critical supercurrent subject to an external
magnetic field in ferromagnetic Josephson junctions. Employing the
quasiclassical Keldysh-Usadel method, we demonstrate theoretically
that an inhomogeneity in the magnitude of the energy
scales in the system, including Thouless energy, exchange field and temperature gradient
normal to the transport direction, influences drastically the
standard Fraunhofer pattern. The exotic non-Fraunhofer response,
similar to that observed in recent experiments, is described in terms
of an intricate interplay between multiple '$0$-$\pi$'-states
and is related to the appearance of
proximity vortices.
\end{abstract}


\maketitle

The well-known Fraunhofer diffraction pattern of the critical
Josephson current has been extensively studied in
superconductor/normal-metal/superconductor (S/N/S) junctions
\cite{cite:clarke,cite:Barzykin}. The interest in how a supercurrent
responds to an applied magnetic flux derives from the fact that this
property is the key element in ultra-sensitive devices such as
superconducting quantum interference devices (SQUID)
\cite{cite:ryazanov1,cite:weides,cite:goldobin}. Whereas S/N/S
junctions are known to display Fraunhofer diffraction in the wide
junction limit, the critical current decays monotonically as a
function of the applied flux in the narrow junction limit. The
crossover between these two distinct types of behavior was
theoretically described in terms of proximity vortices in the normal
wire \cite{cite:bergeret1}.

 More recently, the orbital response of
the supercurrent in magnetic Josephson junctions has attracted much
interest \cite{cite:ryazanov1,cite:weides}. When the normal-metal
interlayer is exchanged with a ferromagnet, thus forming an \sfs
junction, a new mechanism comes into play compared to the \sns case.
The ground-state phase difference between the superconducting
reservoirs may then take the value of $0$ or $\pi$, depending on
parameters such as temperature and ferromagnetic barrier thickness
\cite{cite:ryazanov2}. Not only does this cause the supercurrent in
magnetic Josephson junctions to decay in a non-monotonic fashion,
but it was recently reported that non-Fraunhofer interference
patterns appear in \sfs junctions composed of alternating $0$- and
$\pi$-states \cite{cite:ryazanov1,cite:weides}. Whereas the
supercurrent is maximal at zero flux in the non-magnetic case, the
supercurrent instead displayed a minimum at zero flux in the \sfs
case \cite{cite:weides}. These
experimental findings have motivated theoretical
investigations \cite{cite:goldobin, cite:goldobin2}. Non-Fraunhofer interference patterns have also
been studied in S/I/S junctions with arrays of resistors
\cite{cite:Itzler}.

\begin{figure}[t!]
\includegraphics[width=6.5cm,height=5cm]{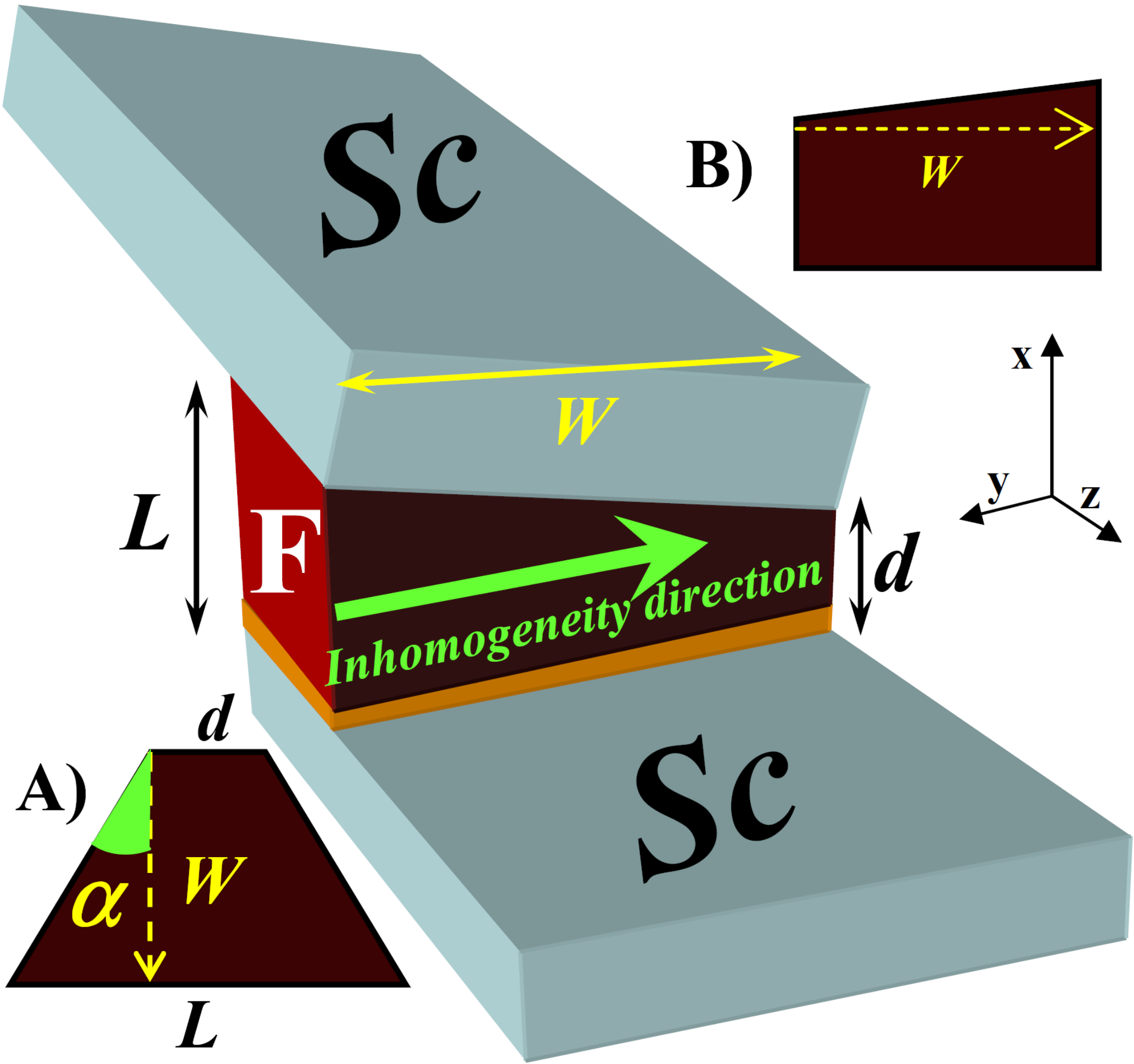}
\caption{\label{fig:model} Experimental setup of the inhomogeneous
ferromagnetic Josephson junction. The green arrow shows the direction of the
inhomogeneity in magnitude of either the Thouless energy, magnetic
exchange field or temperature normal to transport direction
($x$-direction). To model an inhomogeneity in the Thouless energy, the setup
\textbf{A)} is considered in this letter. Although making the set up
\textbf{B)} might be easier to achieve technically, the two setups
generate the same results in the diffusive limit.
An inhomogeneity in the magnetic exchange field is modeled by
$\textbf{h}=\text{h}(0,0,y/W)$. The external field is applied to the
system in the $z$-direction (not shown).}
\end{figure}

Motivated by this, the following question is answered in this
Letter: under which general conditions may the critical supercurrent
respond to an external magnetic field in an anomalous fashion,
producing a non-Fraunhofer interference pattern? We establish these
conditions and moreover explain the origin of this exotic
phenomenon. To do so, we solve the quasiclassical Keldysh-Usadel
equations. In the majority of past works, the investigation of the
non-Fraunhofer patterns were restricted to incorporating a linear
\textit{ansatz} for the behavior of the superconducting U(1) phase
\cite{cite:ryazanov1,cite:weides}. In contrast, we have employed in
this Letter a model of a ferromagnetic Josephson junction which
takes into account an external magnetic field with an arbitrary
dependence on the coordinates and direction of the field. This model
allows us to study the critical supercurrent through an
inhomogeneous junction \textit{without recourse to any ansatz}. The
possibility of having an arbitrary inhomogeneous magnetization
texture in the F region makes the model highly general. The results
of the developed theory are qualitatively in agreement with the
recent experimentally observed non-Fraunhofer patterns. Remarkably,
we find that the critical Josephson current through the F region is
suppressed at zero-external magnetic field within the wide junction
limit when the magnitude of any of the energy scales of the system,
\ie Thouless energy, exchange field and temperature are
inhomogeneous normal to the transport direction. Crucially, to
achieve a non-Fraunhofer response, the inhomogeneity must include at
least one $0$-$\pi$-state. In this case, the second peak of magnetic
interference pattern of critical supercurrent becomes larger than
the first, in contrast to the Fraunhofer pattern. We explain this
behavior in terms of $0$-$\pi$ crossover states and also relate our
results to the appearance of proximity vortices inside the F region.

Consider the schematic of the proposed experimental setup in Fig.
\ref{fig:model}. The inherent Josephson penetration depth
$\lambda_J$ is assumed to be larger than the width of junction, such
that one may avoid screening effects imposed by the Josephson
current on the external magnetic field. This field is assumed to be
directed along the $z$-direction. We work with a vector potential
satisfying the Lorentz gauge \ie
$\overrightarrow{\nabla}$$\cdot$$\mathbf{A}$=$0$ and choose
specifically $\mathbf{A}$=$-Hy\hat{x}$ in which $H$ represents the
strength of external magnetic field. The magnetic flux due to the
intrinsic magnetization of the ferromagnetic region is ignored, as
is known to be a good approximation in most cases
\cite{cite:weides}. To investigate the transport
properties of this system, the quasiclassical theory of
superconductivity in the diffusive regime is employed, so that the Gor'kov equations are reduced to the
Usadel equations \cite{cite:usadel}. The Usadel equation inside
the F region together with appropriate boundary conditions is used
for obtaining observable quantities of the system. In the presence
of a static external magnetic field, the Usadel equation is
succinctly given by;
\begin{eqnarray}\label{usadel}
&&D[\hat{\partial},\check{\text{g}}(x,y,z)[\hat{\partial},\check{\text{g}}(x,y,z)]]+i[
\varepsilon \hat{\rho}_{3}+\nonumber\\&&
\text{diag}[\textbf{h}(x,y,z)\cdot\underline\sigma,(\textbf{h}(x,y,z)\cdot\underline\sigma)^{\tau}],\check{\text{g}}(x,y,z)]=0,
\end{eqnarray}
where $\mathbf{h}(x,y,z)$ stands for exchange energy,
$\check{\text{g}}$ is the full $8\times8$ Green's function matrix,
while $\hat{\rho}_{3}$ and $\underline{\sigma}$ are $4\times 4$ and
$2\times 2$ Pauli matrixes, respectively \cite{cite:morten}. Here
$D$ is the diffusion constant
 and $\hat{\partial}\equiv\overrightarrow{\nabla} \hat{1}-ie \mathbf{A}(x,y,z)\hat{\rho}_{3}
 $. Within the weak proximity regime, one may can expand the Green function
around the bulk solution $\hat{\text{g}}_{0}$ \ie
$\hat{\text{g}}(x,y,z)\simeq\hat{\text{g}}_{0}+\hat{f}(x,y,z)$,
where $\hat{\text{g}}_{0}=\text{diag(\underline{1},-\underline{1})}$
\cite{cite:bergeret2}. Therefore, the retarded component of Green's
function reads:
\begin{align}
\hat{\text{g}}^{R}(x,y,z)\approx\begin{pmatrix}
\underline{1} & \underline{f}^{R}(x,y,z)\\
-\underline{\tilde{f}}^{R}(x,y,z) & -\underline{1}\\
\end{pmatrix}.
\end{align}
The advanced and Keldysh blocks are also given via
$\hat{\text{g}}^{A}(x,y,z)$=-$(\hat{\rho}_3$${\hat{\text{g}}^{R}}(x,y,z)$$\hat{\rho}_3)^{\dag}$
and
$\hat{\text{g}}^{K}(x,y,z)$=$(\hat{\text{g}}^{R}(x,y,z)$-$\hat{\text{g}}^{A}(x,y,z))$$\text{tanh}(\varepsilon/2k_BT)$
under equilibrium conditions. At the two N/S interfaces the
Kupriyanov-Lukichev boundary conditions \cite{cite:zaitsev} are
compactly written by: $2\zeta \hat{\text{g}}
[(\overrightarrow{\nabla} -ie \mathbf{A}\hat{\rho}_{3})\cdot\hat{n}
 ,\hat{\text{g}}] =
[\hat{\text{g}}_\text{BCS}(\phi), \hat{\text{g}}]$, the ratio
between the resistance of the barrier region and the resistance in
the F film is defined as $\zeta$=$R_B/R_F$,
$\hat{\text{g}}_\text{BCS}(\phi)$ is the Green's function in the two
superconductor reservoirs \cite{cite:alidoust} and $\hat{n}$ is a
unit vector normal to the interface \cite{cite:bergeret2}. At the
vacuum borders, the Green's function satisfies
$\partial_y\hat{\text{g}}=0$.

The Usadel equations in their present form constitue a set of complicated
coupled differential equations which we have solved
numerically by using a collocation method. Thus, the approximate
solution components are assumed to be linear combinations of bicubic
(tricubic, for three-dimensional problems) Hermite basis functions,
and required to satisfy the Usadel equation exactly at 4 (8, for
three-dimensional problems) collocation points in each subrectangle
of a grid, and to satisfy the boundary conditions exactly at certain
boundary collocation points \cite{cite:sewell1}. Finally, Newton's
method is used to solve the (nonlinear, generally) algebraic
equations resulting from the collocation method formulation
\cite{cite:sewell3}. In order to study transport properties of the
inhomogeneous junction, the current density through the junction is
considered:
$\mathbf{J}\text{(}\overrightarrow{R},\phi\text{)}=J_{0}\int
d\varepsilon\text{Tr}\{\rho_{3}(\check{\text{g}}[\hat{\partial},\check{\text{g}}])^{K}\}$,
here $J_{0}=N_{0}eD/4$ and $N_{0}$ is the number of states in the
Fermi surface. Performing an integration over the $y$-coordinate
provides the total supercurrent flowing through the junction,
$I\text{(}\phi\text{)}=I_{0}\int\int
dyd\varepsilon\text{Tr}\{\rho_{3}(\check{\text{g}}[\hat{\partial},\check{\text{g}}])^{K}\}$.
To understand the magnetic interference patterns of such junctions,
we also investigate the spatial variation of pair potential inside
the F region calculated via:
$U=U_{0}\text{Tr}\{(\hat{\rho}_{1}-i\hat{\rho}_{2})\int d\varepsilon
\hat{\tau}_{3}\check{\text{g}}^{K}\},$ where
$U_{0}=-N_{0}\lambda/16$ \cite{cite:morten}.
The temperature, width and lower base of the \textit{wedged} junction
are fixed at $T/T_c=0.01$, $W/\xi_S=10$ and $d/\xi_S=2$,
respectively (the so-called wide junction limit). The proximity
controlling parameter $\zeta$ is also fixed at $5$, ensuring
that we operate in the weak proximity regime. Energy units are used
so that $\hbar$=$k_B$=1.

\begin{figure*}[t!]
\includegraphics[width=14.5cm,height=5.5cm]{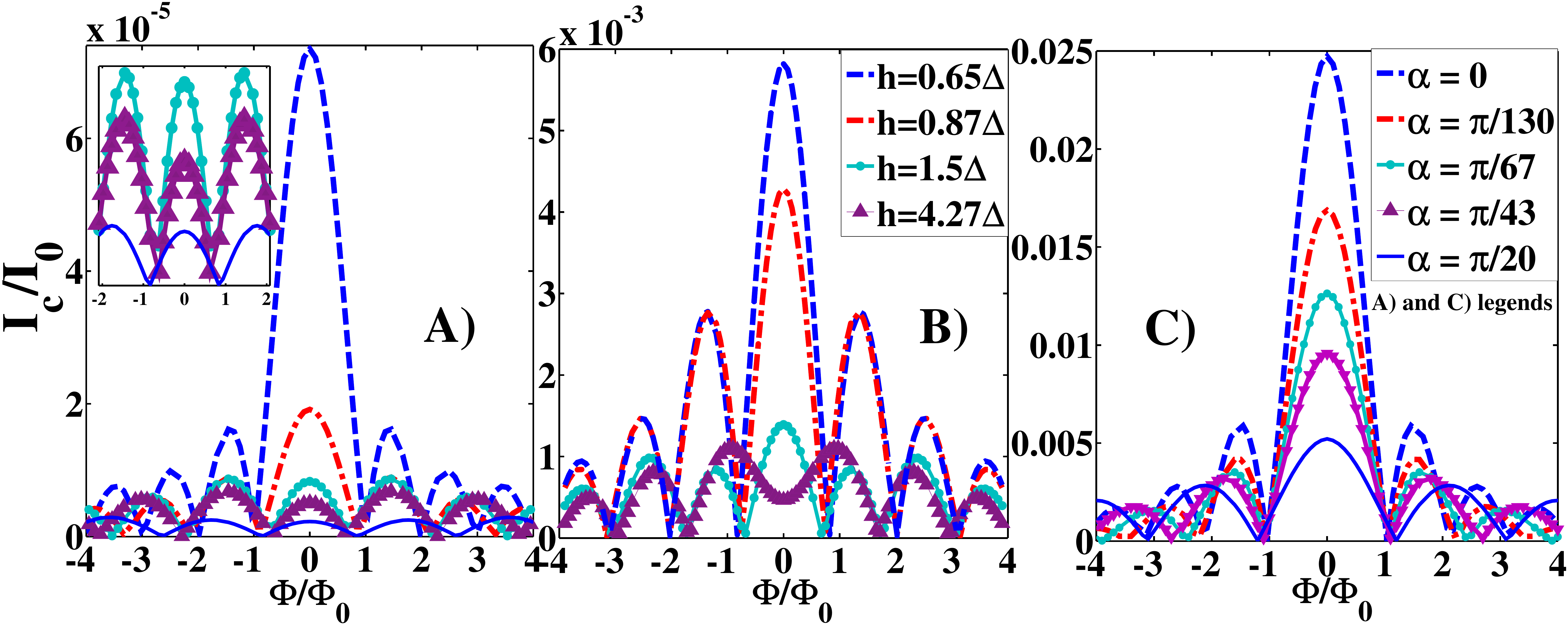}
\caption{\label{fig:NonFraunhofer} Normalized supercurrent through
an inhomogeneous Josephson junction vs normalized external magnetic
flux $\Phi/\Phi_0$ perpendicular to the junction. \textbf{A)}: An
\sfs junction where the F region has a wedged shape with inclination
angle $\alpha$ and exchange field $h=10\Delta_0$. The inset panel
zooms in on the interference pattern at small and large values of
$\Phi$ and $\alpha$, respectively. \textbf{B)}: The F region is now
geometrically rectangular, but the magnitude of exchange field is
now inhomogeneous according to the texture
$\textbf{h}=\text{h}(0,0,y/W)$. \textbf{C)}: An \sns junction where
the N region has a wedged shape with inclination angle $\alpha$. The
legends are the same as in \textbf{A)}. In all cases, the height of
the trapezoidal region is fixed at $W=10\xi_S$ while the upper base
$d$ is equal to $2\xi_S$ (see Fig. \ref{fig:model}). Therefore,
$\alpha$=0 makes a rectangular junction ($L=d=2\xi_S$) and $\pi$/130
($L = 2.483\xi_S$), $\pi$/67 ($L = 2.938\xi_S$), $\pi$/43 ($L =
3.463\xi_S$) and $\pi$/20 ($L = 5.168\xi_S$) make wedged junctions.
}
\end{figure*}

The results for the critical Josephson current through the
inhomogeneous \sfs junction as a function of normalized external
magnetic flux \ie $\Phi/\Phi_0$ are presented in Fig.
\ref{fig:NonFraunhofer}. In frame \textbf{A)}, the magnitude of the
Thouless energy is inhomogeneous in the $y$-direction: the F region
has a wedged shape [see \textbf{A)} in Fig.
\ref{fig:model}]. In frame $\textbf{B)}$, the magnitude of magnetic
exchange interaction is inhomogeneous in the $y$-directions and
follows a $\textbf{h}=\text{h}(0,0,y/W)$ pattern. The normalized
critical current $I_c$/$I_0$ exhibits a suppression at zero external
flux for some values of the wedge angle $\alpha$. In the case of
trapezoidal junction, the second peak in the interference pattern
takes a larger value than the first for an interval of
$\alpha$-values. Typically, this behavior is enhanced at
$\alpha$=$\pi/43$ and then disappears for larger values
$\alpha>\pi/20$ (see inset panel of panel \textbf{A)} of Fig.
\ref{fig:NonFraunhofer}). A similar magnetic interference pattern is
generated when the magnitude of exchange field is inhomogeneous, as
shown in frame \textbf{B)}. In this case, the non-Fraunhofer
pattern phenomenon is pronounced for \eg $h=4.27\Delta$. The results
show qualitatively good consistency with recently reported
non-Fraunhofer patterns for "$0$-$\pi$"-stacks in Ref.
\onlinecite{cite:weides}. For reasons to be described below, we
expect that the same non-Fraunhofer magnetic pattern would arise
when the temperature of the system along the $y$-direction is
variable and has an inhomogeneous form.
We have also investigated (not
shown) other magnetization textures such as domain-wall, skyrmion
and spiral (with helical axis normal to transport direction), and found that they
generate the standard Fraunhofer patterns because of the
constant magnitude of the magnetic exchange field,
$|$\textbf{h}$|$=h.

In comparison, the behavior of critical Josephson current through a
\sns wedged junction is investigated in Fig. \ref{fig:NonFraunhofer} \textbf{C)}.
For $\alpha=0$, we recover the results of Ref.
\onlinecite{cite:bergeret1}. With increasing $\alpha$, the
normalized supercurrent is subject to an overall reduction, because
the effective junction length increases due to the inhomogeneity in
the magnitude of Thouless energy. Unlike the inhomogeneous \sfs case
above, however, the first peak in the diffraction pattern is larger
than others for all values of $\alpha$, exhibiting the standard
Fraunhofer pattern. Therefore, the exotic non-Fraunhofer pattern only can
be observed in ferromagnetic junctions under the conditions discussed
above.

\begin{figure*}[t!]
\includegraphics[width=17cm,height=9.5cm]{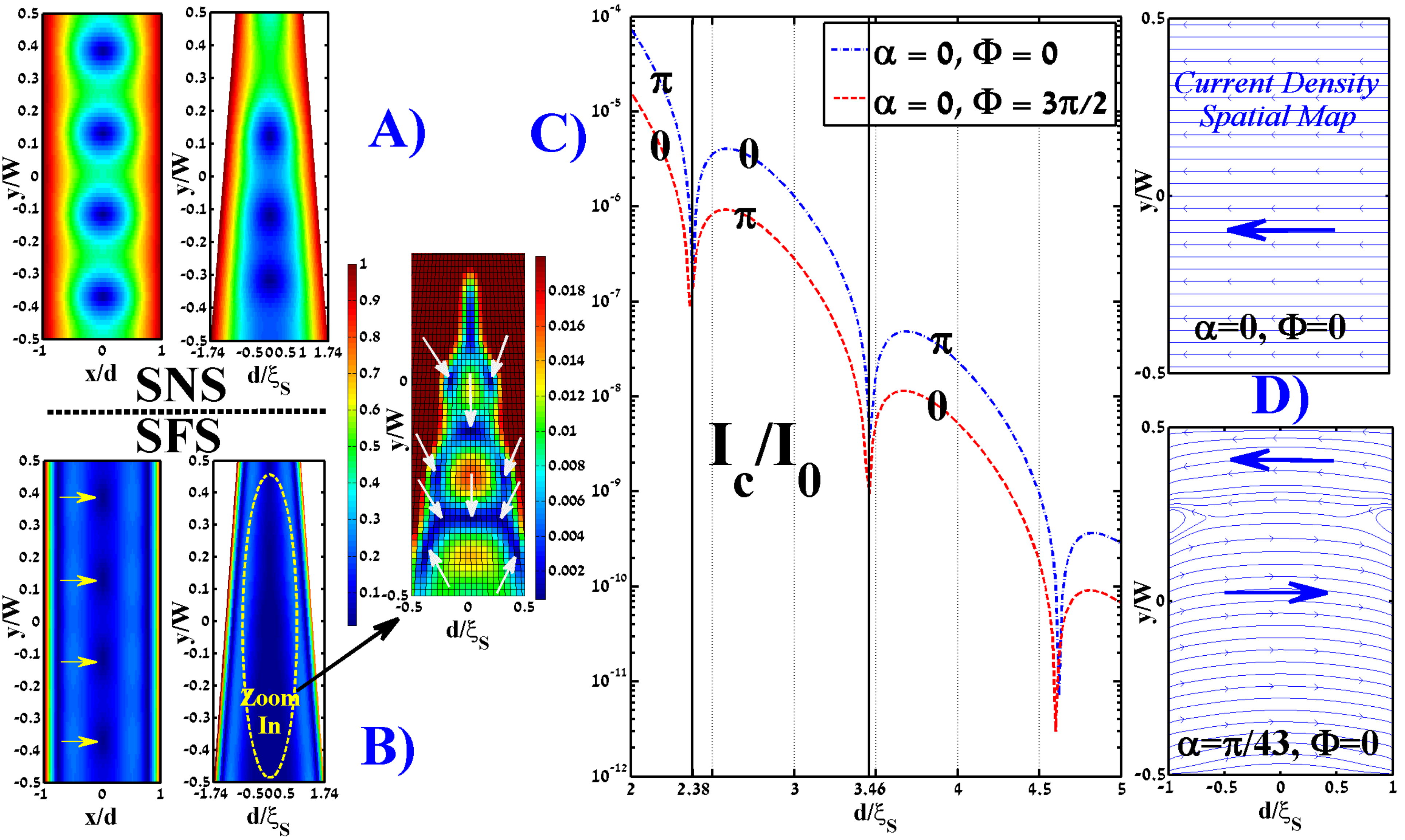}
\caption{\label{fig:crossover} \textbf{A)} Normalized spatial map of
the pair potential for a \sns junction where $\alpha$=$0$, $\pi/43$
($L$=3.463$\xi_S$) in the left and right frames, respectively.
\textbf{B)} The normalized pair potential in a \sfs junction. Arrows
indicate the location of proximity vortices. The zoom-in frame of a
\sfs trapezoidal junction with $\alpha$=$\pi/43$ is shown using a
different color map. The magnetic flux through the N or F region is
assumed to be $\Phi$=$4\Phi_0$ and no superconducting phase
difference is applied ($\phi$=0). \textbf{C)} The $0$-$\pi$
crossover profile of a rectangular \sfs junction where $\Phi$=0,
$3\Phi_0/2$ and $W=10\xi_S$ vs. normalized junction length
$d/\xi_S$. \textbf{D)} Current density spatial map of the magnetic
Josephson junction in the absence of external magnetic field,
$\Phi$=0. Top and bottom frames exhibit current density flowing
through the junction where $\alpha$=$0$, $\pi/43$, respectively
($\alpha$=$\pi/43$ constitutes a '0-$\pi$'-junction). Blue arrows
indicate current directions.}
\end{figure*}

To further understand the outstanding difference between
interference patterns of the homogeneous and inhomogeneous
ferromagnetic junctions, we consider how the presence of an external
magnetic field influences both the 0-$\pi$ transition profile of
the \sfs junction and the proximity vortices pattern in the F
region. The origin of the suppressed central peak in the \sfs
wedged Josephson junction ($\alpha\neq0$) is mainly studied in this
Letter, and we argue why this
mechanism accounts for the non-Fraunhofer pattern in other cases where
there is a magnitude gradient of the exchange field and/or temperature along the direction normal to the transport direction
so that it includes at least one 0-$\pi$-state. To this end, we will later
investigate the current density spatial map of magnetic junctions
and compare $\alpha$=0 with $\alpha\neq0$. Fig. \ref{fig:crossover} reveals an
illustrative profile of the normal and ferromagnetic Josephson
junctions. Part \textbf{A)} illustrates a spatial map of pair
potential in the normal region of the \sns junction where
$\alpha$=$0$ and $\pi/43$ for the left and right panels,
respectively. The upper base of the trapezoidal region is fixed at
$d=2\xi_S$ while for $\alpha$=$\pi/43$ the lower base takes the
value $L$=$3.463\xi_S$. The increment of $\alpha$ deforms the
proximity vortex pattern compared to the pattern of the rectangular
junction. The distance between two neighboring vortices is no longer
equal to $\Phi_0 H/d$ in contrast to that of rectangular junction
within the wide junction regime (left panel). The spatial maps of
the pair potential are given for $\Phi$=$4\Phi_0$ and zero
superconducting phase difference \ie $\phi$=0. Now, the increment of
$\alpha$ removes the proximity vortices inside the normal segment of
the junction gradually. The variation of superconducting phase
difference $\phi$ however, moves the vortices along the
$\hat{y}$-direction in both the normal and ferromagnetic junctions
\cite{cite:bergeret1}. Part \textbf{B)} exhibits the equivalent
investigation for a ferromagnetic Josephson junction. In the
rectangular case, the pair potential shows the same behavior as the
vortex pattern as \sns case. In contrast, the vortex pattern is
highly deformed in the wedged ferromagnetic junction. A zoom-in is
shown for the middle of F wire with $\alpha$=$\pi/43$ using
different color map. The strong deformation may be understood by
noting that the increment of $\alpha$ effectively synthesizes
multiple "$0$-$\pi$" states in the same junction and also from the pair-breaking of the exchange field.
When $\alpha$
becomes non-zero, the junction may be thought of as a superposition
of multiple 0 and $\pi$ junctions.

To understand this quantitatively,
parts \textbf{C)} and \textbf{D)}
should be considered together. Part \textbf{C)} illustrates the
0-$\pi$ crossover profile where $\alpha=0$ for two different values
of external flux $\Phi$=$0$, $3\Phi_0/2$ as a function of F-layer
length $d/\xi_S$. The first and second transitions occur at
$d$=$2.38\xi_S$ and $3.46\xi_S$, respectively. It is important to
note that the latter length is identical to the lower base of the
trapezoidal junction when $\alpha$=$\pi/43$. The plot also shows
that applying an external magnetic field reduces the magnitude of
the current nonlinearly, although the locations of $0$-$\pi$ points
are left unchanged. The anomalous
Fraunhofer diffraction pattern of the critical supercurrent can now
be well understood by noting part \textbf{D)}. Increasing the
junction angle $\alpha$ renders more parts of the junction to have
opposite supercurrent flow direction which then partially cancel
each other. One should note that in the trapezoidal region, the
amplitude of critical current is non-uniform. More $\pi$-parts,
therefore, are needed to cancel the $0$-parts of the junction that
occur for the top region with smallest effective length $L$. A key
observation is that the above results suggest a venue for producing
an anomalous non-Fraunhofer interference pattern resulting without
necessarily distorting the geometry of the system, in effect
allowing for anomalous interference even in standard rectangular
junctions.

In conclusion, we have proposed experimentally accessible generic
conditions for achieving non-Fraunhofer interference patterns of the
critical supercurrent as a function of external magnetic flux. The
key property is the controllable numbers of gradual $0$-$\pi$ states
in the same junction by incorporating an inhomogeneity in the
magnitude of energy scales of system \ie Thouless energy, exchange
field and/or temperature normal to the transport direction. We examine
the proposed generic conditions for some limiting cases and find
good qualitative consistency with the recently observed
non-Fraunhofer magnetic interference patterns in $0$-$\pi$ stacks.

\acknowledgments

The authors are grateful to F.
S. Bergeret and J. W. A. Robinson for valuable discussions, and thank A. Sudb${\o}$ for conversations during the initial stages of
this work and K. Halterman for his generosity regarding compiler
source.


\begin{thebibliography}{000}
\bibitem{cite:sewell1} Although the collocation method seems to be
applicable only to rectangular regions, it can actually be used to
easily solve problems in a wide range of simple two-dimensional and
three-dimensional regions, as is outlined in Ref.
\onlinecite{cite:sewell2b}.
\end{thebibliography}
\end{document}